\documentclass[aps,floats,floatfix,amssymb,amsmath,prb,nofootinbib,twocolumn,superscriptaddress]{revtex4-1}

\usepackage{calc}
\usepackage{psfrag}
\usepackage{graphicx}
\usepackage{color}
\usepackage[dvipsnames]{xcolor}
\usepackage[unicode=true,pdfusetitle,
 bookmarks=true,bookmarksnumbered=false,bookmarksopen=false,
 breaklinks=false,pdfborder={0 0 0},backref=false,colorlinks=true]
 {hyperref}
\usepackage{geometry}
\usepackage[normalem]{ulem}

\pdfpageattr {/Group << /S /Transparency /I true /CS /DeviceRGB>>} 

\usepackage{pdfpages}

\makeatletter
\AtBeginDocument{\let\LS@rot\@undefined}
\makeatother

\begin{document}

\title{Conductivity in the square lattice Hubbard model at high temperatures: importance of vertex corrections}
\author{J. Vu\v ci\v cevi\'c}
\affiliation{Scientific Computing Laboratory, Center for the Study of Complex Systems, Institute of Physics Belgrade,
University of Belgrade, Pregrevica 118, 11080 Belgrade, Serbia}
\author{J. Kokalj}
\affiliation{University of Ljubljana, Faculty of Civil and Geodetic Engineering,  Jamova 2, Ljubljana, Slovenia}
\affiliation{Jozef Stefan Institute, Jamova 39, SI-1000, Ljubljana, Slovenia}
\author{R. \v Zitko}
\affiliation{Jozef Stefan Institute, Jamova 39, SI-1000, Ljubljana, Slovenia}
\affiliation{University of Ljubljana, Faculty of Mathematics and Physics,  Jadranska 19, Ljubljana, Slovenia}
\author{N. Wentzell}
\affiliation{Center for Computational Quantum Physics, Simons Foundation Flatiron Institute, New York, NY 10010 USA}
\author{D. Tanaskovi\'c}
\affiliation{Scientific Computing Laboratory, Center for the Study of Complex Systems, Institute of Physics Belgrade,
University of Belgrade, Pregrevica 118, 11080 Belgrade, Serbia}
\author{J. Mravlje}
\affiliation{Jozef Stefan Institute, Jamova 39, SI-1000, Ljubljana, Slovenia}

\begin{abstract}
Recent experiments on cold atoms in optical lattices allow for a quantitative comparison of the measurements to the conductivity calculations in the square lattice Hubbard model. However, the available calculations do not give consistent results and the question of the exact solution for the conductivity in the Hubbard model remained open. In this letter we employ several complementary state-of-the-art numerical methods to disentangle various contributions to conductivity, and identify the best available result to be compared to experiment. We find that at relevant (high) temperatures, the self-energy is practically local, yet the vertex corrections remain rather important, contrary to expectations. The finite-size effects are small even at the lattice size $4\times 4$ and the corresponding Lanczos diagonalization result is therefore close to the exact result in the thermodynamic limit.
\end{abstract}
\maketitle

Theoretical study of transport in condensed matter systems with strong interactions is very difficult.
In many cases there are no long-lived quasi-particles, and the conventional Boltzmann theory of transport provides little insight.
Progress can only be made using bona fide many-body approaches to simplified lattice models or effective field theories,
where approximations are made in a controlled manner.~\cite{terletska11,deng13,xu13,Vucicevic2015,pakhira15,perepelitsky16,kokalj17,huang18,hartnoll_book}
Even then, as only a few specifics of a real system enter the model, the
comparison to relevant experiments can only be made at a qualitative
level.
This changed very recently, when Ref.~\onlinecite{brown18} reported a measurement of transport in a quantum simulator of the fermionic Hubbard model in two dimensions (2D). 
The experiment is performed on cold lithium atoms in an optical lattice, a controllable setup free from disorder, phonons and other complications of realistic materials.
It is well justified to compare \emph{at the quantitative level} such experimental result for conductivity with the Hubbard model calculations.

Ref.~\onlinecite{brown18} found that two state-of-the-art methods namely the
finite-temperature Lanczos method (FTLM) and the dynamical mean field theory (DMFT) give conductivities that differ 
by up to a factor $\frac{3}{2}$, and only FTLM shows a solid agreement
with the experiment. At high temperatures $T \gtrsim t$  relevant to
these observations (for instance, in cuprates where
the  hopping parameter $t\approx 0.3$eV the corresponding temperature is well above the melting temperature) one expects the correlation lengths to be short, and
the approximations made in the two methods to apply.  Our aim is to
reveal the physical origin of this discrepancy and to establish a
numerically exact solution in the regime $T/t\gtrsim 1$ relevant for
optical lattice experiments, as well as other narrow band systems,
such as organic superconductors~\cite{powell11}, low temperature phase
of TaS$_2$~\cite{rossnagel06}, twisted bilayer graphene~\cite{cao18},
and monolayer transition metal dichalcogenides~\cite{coleman11}, such
as 1T-NbSe$_2$~\cite{nakata16}.

\begin{figure}[!t]
 \includegraphics[width=3.2in,trim=0cm 0cm 0cm 0cm]{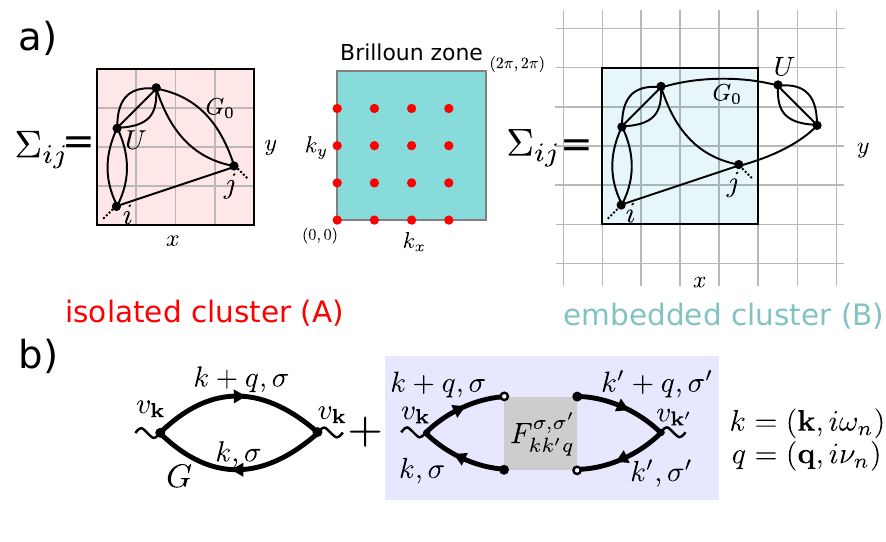}
 \caption{
 (a) Illustration of the type of self-energy diagrams that are captured by isolated cluster and embedded cluster (in particular cellular DMFT), and the respective difference in the Brillouin zone (discrete vs. continuous). 
 (b) Separation of a susceptibility into the bubble and the vertex corrections part. 
}
\label{fig:illustration} 
\end{figure}

It is useful to recall that the mentioned
numerical methods belong to two distinct general approaches: $A$) one solves
an isolated finite cluster of lattice sites, as representative of the
thermodynamic limit;~\cite{Trivedi1996,kokalj17,huang18} $B$) one solves an effective, self-consistently
determined ``embedded'' cluster, which provides propagators of
infinite range, yet limits the range of electronic correlations.~\cite{Biroli2002,Maier2005a,Kotliar2006,GullRMP2011,Leblanc2015,Ayral2016,Ayral2017a,Vucicevic2018,Rohringer2018} The diagrammatic content of the self-energy in the two approaches is
sketched in Fig.~\ref{fig:illustration}a.
Approach $B$ captures longer distance quantum fluctuations, and is therefore assumed to converge more quickly with cluster size
at the price of an iterative solution of the (embedded) cluster, as opposed to the ``single-shot'' calculation in the approach $A$.
FTLM solves a $4\times 4$ isolated cyclic cluster and belongs to $A$.
DMFT is an embedded cluster calculation ($B$) with the cluster size one, and therefore it approximates the self-energy by a purely local quantity.

Therefore, there are three possible sources of discrepancy between the DMFT and FTLM results for resistivity:
(i) {\it non-local correlations} which are encoded in the non-local corrections to self-energy, present in FTLM but beyond the DMFT approximation;
(ii) {\it quantum fluctuations} at distances beyond the linear size of the FTLM cluster; DMFT captures them through an effective fermionic bath;
(iii) {\it vertex corrections}, included within FTLM, but neglected within DMFT where one calculates only the bubble contribution . We recall that the
two-particle correlation functions can be split into
the disconnected part (``the bubble'') and the connected part (``vertex corrections''), as shown in Fig.~\ref{fig:illustration}b.
The bubble captures only the single-particle scattering off the medium, described by the self-energy which enters the full Green's function.
The collective excitations come from the particle-hole scattering, and are present only in the vertex corrections.
Whereas the contribution of the connected part is always important for charge
susceptibility 
~\cite{hafermann14,nourafkan18,krien18}, in the large dimensionality limit the
vertex corrections to conductivity cancel ~\cite{khurana90} (the full vertex $F$ loses $\mathbf{k}\mathbf{k}'$-dependence and the current vertex is odd $v_{-\mathbf{k}}=-v_\mathbf{k}$, unlike the charge vertex which is even).
In finite
dimensions, however, the vertex corrections do contribute to conductivity, as discussed previously in several approximative approaches at low temperatures~\cite{lin09,lin10,bergeron11,sato12,lin12,sato16,Kauch2019}. Based on the Ward identity one could think that
when the correlations are approximately local, the vertex corrections become negligible \cite{bergeron11,lin09}.
We show that this expectation is not satisfied\cite{comment3},
and that despite the non-local self-energy being practically negligible at $T \gtrsim 0.3D$, the vertex corrections still amount for a sizable shift in dc-resistivity.
Additionally, we show that long-distance quantum fluctuations have little effect on dc conductivity, thus rendering a $4\times 4$ isolated-cluster calculation sufficient to obtain exact results for the bulk model.

\emph{Model.}
We consider the Hubbard model on the square lattice
\begin{equation}
 H = -t\sum_{\sigma, \langle i,j\rangle} c^\dagger_{\sigma i} c_{\sigma j}+U\sum_{i} n_{\uparrow i}n_{\downarrow i}-\mu\sum_{\sigma, i} n_{\sigma i},
\end{equation}
where $c^\dagger_{\sigma i}/c_{\sigma i}$ create/annihilate an electron of spin $\sigma$ at the lattice site $i$. The hopping amplitude between the nearest neighbors is denoted $t$, and we set $D=4t$ as the unit of energy. 
We also take lattice spacing $a=1$, and $\hbar=e=1$.
The density operator is $n_{\sigma i}=c^\dagger_{\sigma i}c_{\sigma i}$, the chemical potential $\mu$, and the on-site Hubbard interaction $U$. Throughout the paper, we keep $U=2.5 D$, which corresponds to the (doped) Mott insulator regime, and assume paramagnetic solutions with full lattice symmetry.

\begin{figure*}[!ht]
 \includegraphics[width=6.4in,trim=0cm 0cm 0cm 0cm]{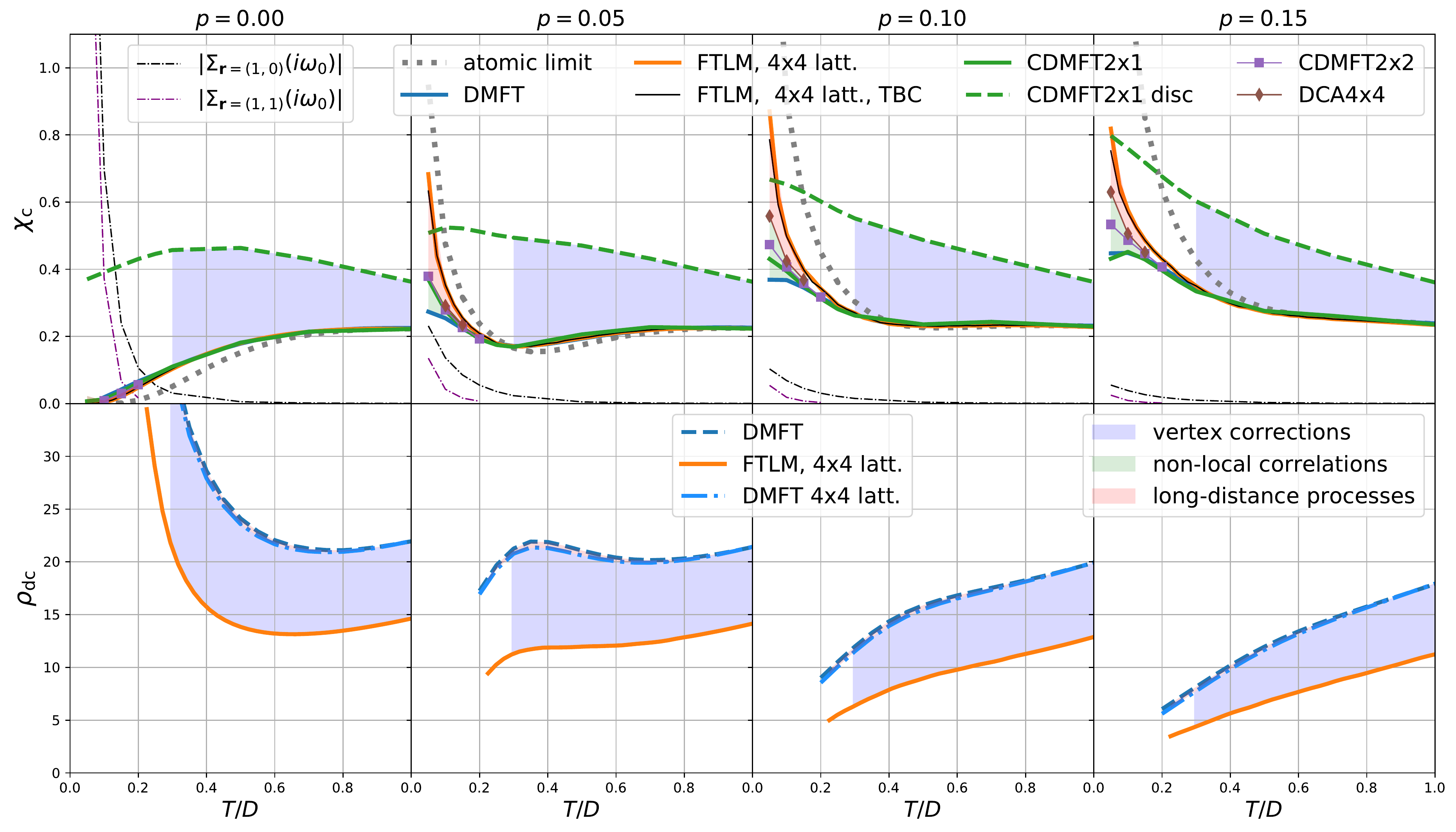}
 \caption{
 Charge susceptibility (upper) and dc resistivity (lower) as a function of temperature, at different levels of doping. 
 The color between the curves denotes the physical origin of the difference. Dashed curves denote just the bubble contribution, solid lines the full result.
}
\label{fig:chi_rho} 
\end{figure*}

\emph{Formalism.} 
The conductivity is defined in terms of the current-current correlation function
\begin{equation}
  \Lambda^{xx}_\mathbf{q}(i\nu_n) \equiv \sum_i e^{-i\mathbf{q}\cdot\mathbf{r}_i} \int d\tau e^{i\nu_n\tau} \langle j^x_i(\tau) j^x_{i=0}(0) \rangle ,
\end{equation}
where $\tau$ is imaginary time,  $i\nu_n=2in\pi T$ is bosonic Matsubara frequency, 
$\mathbf{r}_i=(x_i,y_i)$ denotes the real-space vector of the site $i$.
The  current operator $j$ is defined as 
$ j^x_i = -it\sum_\sigma c^\dagger_{\sigma i} c_{\sigma, \mathrm{n.n.}(i; x)} + \mathrm{h.c.}$
where $\mathrm{n.n.}(i; x)$ denotes the nearest neighbor in the $x$ direction.
We are interested in longitudinal, uniform conductivity $\sigma^{xx}_{\mathbf{q}=0}(\omega)$, so we adopt a shorthand notation $\Lambda(i\nu_n) \equiv \Lambda^{xx}_{\mathbf{q}=0}(i\nu_n)$ and $\sigma(\omega) \equiv \sigma^{xx}_{\mathbf{q}=0}(\omega)$.
The optical conductivity is given by\cite{coleman_book} 
$\sigma(\omega) = -(i/\omega)\left[ \Lambda(\omega) - \Lambda(\omega=0)\right] $,
where $\Lambda(\omega)$ is the analytical continuation of $\Lambda(i\nu_n)$ to the real axis, i.e. the inverse of the Hilbert transform

\begin{equation}\label{eq:hilbert}
\Lambda(i\nu) = \frac{1}{\pi}\int d\omega \frac{\mathrm{Im}\Lambda(\omega)}{\omega-i\nu} = \frac{1}{\pi}\int d\omega \frac{\omega\mathrm{Re}\sigma(\omega)}{\omega-i\nu}.
\end{equation}
The second equality in Eq.~\eqref{eq:hilbert} is due to $\mathrm{Im}\Lambda(\omega=0)=0$.
The direct-current (dc) conductivity is defined as $\sigma_\mathrm{dc} = \mathrm{Re}\sigma(\omega=0) = \mathrm{Im}\Lambda'(\omega=0)$, and the dc resistivity is then $\rho_\mathrm{dc}=1/\sigma_\mathrm{dc}$.

In order to better identify and understand the importance of various processes for the transport, we also calculate the charge susceptibility $\chi_\mathrm{c} = d \langle n \rangle / d\mu $, which corresponds to the charge-charge correlation function\cite{comment4}.
Both $\chi_\mathrm{c}$ and $\Lambda$ and can be separated into the bubble and the vertex corrections part\cite{comment5}, Fig.~\ref{fig:illustration}.
In all quantities, the superscript ``disc'' denotes the bubble contribution, and the superscript ``conn'' the vertex corrections part.

\emph{Methods} $A$.
We solve an isolated cyclic $4\times 4$ cluster
using the FTLM~\cite{jaklic00,jaklic95} method and both $4\times 4$ and $8\times 8$ using quantum Monte
Carlo (the continuous-time interaction-expansion algorithm,
CTINT\cite{Rubtsov2004,GullRMP2011})  Both methods yield numerically exact solutions of the
representative finite-size model. 
In FTLM we calculate $\sigma(\omega)$, while CTINT yields
$\Lambda(i\nu_n)$, as well as the self-energy $\Sigma_{ij}(i\omega_n)$
and the Green's function $G_{ij}(i\omega_n)$\cite{comment2}. Note that both CTINT and FTLM allow for a direct calculation of the full current-current correlation function, and that we need not evaluate the full vertex function $F$ at any stage of the calculation. 

In the isolated cluster calculations one faces several finite-size effects stemming from the finite range of the bare electronic propagator\cite{jaklic95,jaklic00}.
Most importantly, this not only limits the range of electronic correlations, 
but also affects the diagrammatic content of short range correlations: 
diagrams with distant interaction vertices are not captured (Fig.~\ref{fig:illustration}).
One may see this equivalently in the $\mathbf{k}$-space as a discretization of the Brillouin zone,
which affects the internal momentum summations in all self-energy and full vertex diagrams.

\begin{figure}[!t]
 \includegraphics[width=3.2in,trim=0cm 0cm 0cm 0cm]{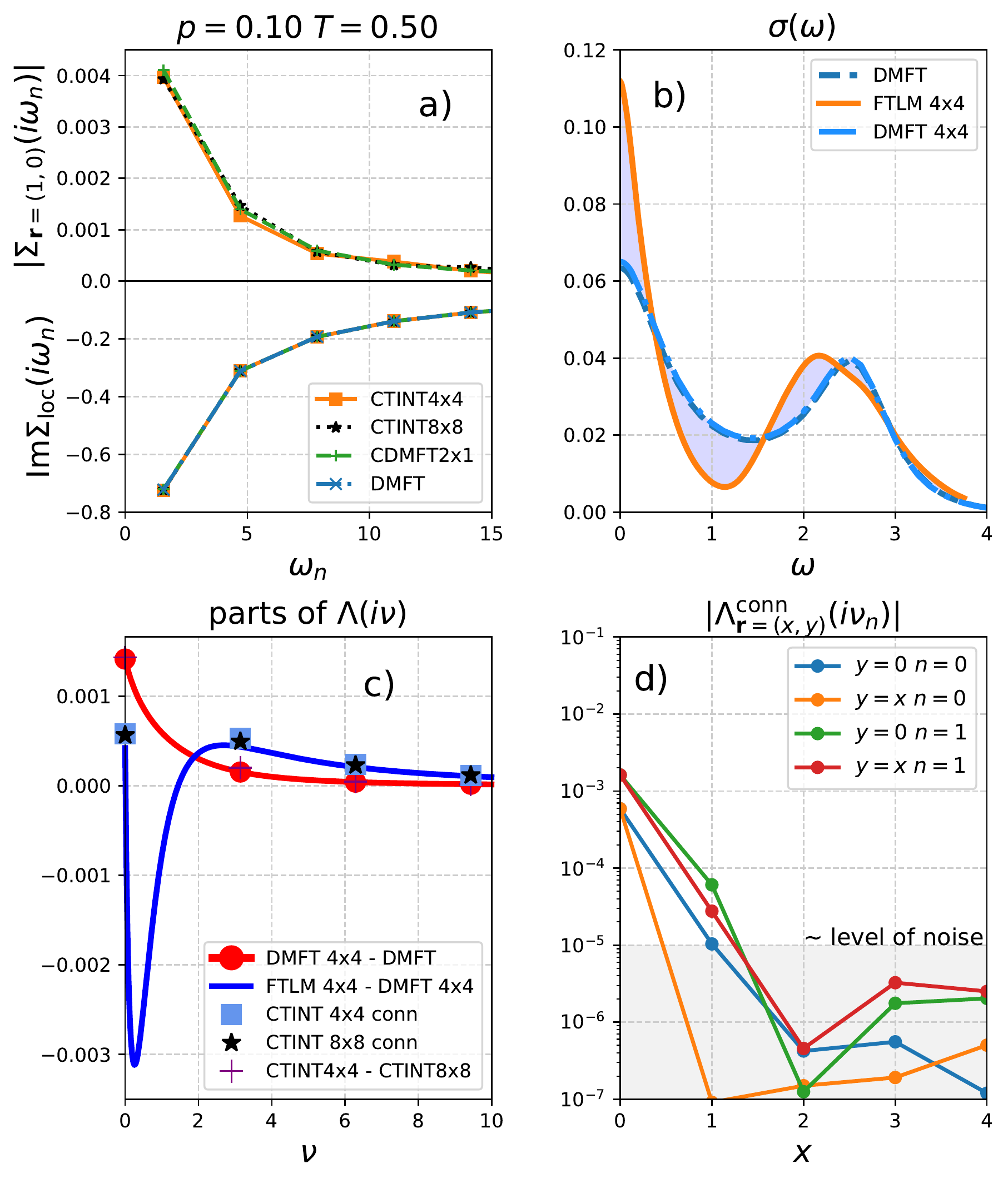}
 \caption{ All panels: $p=0.1$, $T=0.5D$.
 a) Benchmark of self-energy and inspection of its leading non-local component.  
 b) Comparison of the optical conductivity between various methods. 
 c) See text. 
 d) Real-space resolution of the vertex corrections along two spatial directions (CTINT $8\times 8$ result).
}
\label{fig:lambda_sigma} 
\end{figure}

\emph{Methods} $B$.
We solve the embedded clusters of size $2\times 1$ and $2\times 2$ within the cellular DMFT scheme (CDMFT)\cite{Kotliar2001} and the $4\times 4$ cluster within
the dynamical cluster approximation (DCA) scheme\cite{Hettler1998}, both using CTINT. (Unlike the isolated cluster case, the bare propagator entering CTINT here takes into account the effective medium.) 
The single-site DMFT calculations (cluster size
$N_\mathrm{c}=1$) are done using both the CTINT and the approximative real-frequency
numerical renormalization group method (NRG) as impurity solvers. 
 
In CDMFT, an electron can travel infinitely far between two scatterings,
but a self-energy 
insertion in the corresponding diagrammatic
expansion can only be of limited range (see
Fig.~\ref{fig:illustration}).
In DCA, the approximation is made in reciprocal space and amounts to allowing the electron to 
visit $\mathbf{k}$-states otherwise not present in the finite cluster.\cite{Vucicevic2018}

\emph{Results.}
Top panels of Fig.~\ref{fig:chi_rho} show the temperature dependence of
$\chi_\mathrm{c}$ for several values of doping $p=1-\langle
n\rangle$. One sees that in the high-temperature regime $T\gtrsim
0.3D$, the results of different methods (solid curves) all agree and tend toward the atomic limit, as expected for a thermodynamic quantity.

At lower temperatures, the non-local correlations show
up.
Away from half-filling, FTLM and DCA yield a charge susceptibility that increases with lowering temperature, yet in DMFT, it saturates instead.
The enhancement of charge
susceptibility at low $T$ comes from the antiferromagnetic fluctuations~\cite{kokalj17}.
The difference between the DCA
and the DMFT is used to characterize the importance of non-local
correlations (green shading). They manifest themselves
also in the growth of non-local self-energy at low $T$ (thin dashed-dotted lines). 
The DCA and the FTLM result do not completely coincide; the difference (pink shading) comes from
the longer-distance quantum fluctuations.  
The discretization of the Brillouin zone in FTLM can be somewhat ameliorated by the twisted-boundary conditions scheme (TBC)~\cite{poilblanc91}.
As expected, TBC is closer to DCA (black line), but one needs a better method to capture the full effect of longer-range processes.
 
We have also evaluated separately the bubble
contribution $\chi^{\mathrm{disc}}$ to $\chi_c$ (dashed lines)
and observe it is substantially larger than the full result $\chi_c$.
 
Bottom panels of Fig.~\ref{fig:chi_rho} show the temperature dependence
of resistivity $\rho_\mathrm{dc}$ as calculated from the bubble term in the DMFT (dashed line)
and the full result from FTLM (solid line). Strikingly, even in the temperature range
$T\gtrsim 0.3D$ where the behavior of $\chi_c$ collapsed to that
of the atomic limit, the DMFT and FTLM are shown to yield significantly
different results with a lower value of resistivity found in the FTLM.

To understand the origin of this difference we take a closer look of
the data at $T=0.5$, $p=0.1$ that we show in
Fig.~\ref{fig:lambda_sigma}.  In panel a) we compare the self-energies
found in the DMFT, CDMFT 2x1 and the CTINT calculation for the isolated $4\times 4$ and $8\times 8$ clusters. Not only is the
nearest neighbor self-energy (top) found to be two orders of
magnitude smaller than the local one (bottom), but also the
local parts of the self-energies show excellent agreement. Thus,
neither non-local correlations (neglected in DMFT) nor long-range
processes (neglected in $4 \times 4$) play an important role for the
self-energy at this temperature.

Might long-range processes play a more important role for the
conductivity?  
One can readily investigate the role of long-range
processes for the bubble part of the conductivity.
This is done by
calculating the conductivity in the DMFT formulated for the $4\times 4$ lattice,
which amounts to discretizing the Brillouin zone (in both the self-consistency condition, and internal bubble summation (Fig.~\ref{fig:illustration}b).
Fig.~\ref{fig:lambda_sigma}b
compares the optical conductivity obtained in this way (denoted by
DMFT $4\times 4$) to the infinite lattice DMFT result and to the FTLM one. The DMFT and the
DMFT $4\times 4$ are close: the long-range processes clearly do not account for
the discrepancy between the DMFT and the FTLM either.
The most of the
difference between the DMFT and the FTLM conductivity thus comes from the
vertex corrections. 

To further verify this result 
we have evaluated the current-current correlation function $\Lambda(i\nu_n)$ also in CTINT $4\times 4$,
and deduced the connected part by $\Lambda^\mathrm{conn}(i\nu_n)=\Lambda(i \nu_n)-
\Lambda^\mathrm{disc}(i\nu_n)$, which is shown by the blue squares in Fig.~\ref{fig:lambda_sigma}c.
These points fall on the blue line which is obtained by the Hilbert transform to the imaginary axis (Eq.~\ref{eq:hilbert}) of the difference in $\sigma(\omega)$ between the FTLM and the DMFT $4\times 4$ (see  Supplemental Material (SM) for details and other $p, T$). Note that the magnitude of $\Lambda^{\mathrm {conn}}$ at the Matsubara frequencies is rather small, consistent with the Ward identity
 $\Lambda^\mathrm{conn}(i\nu=0) \sim \sum_\mathbf{k} v_\mathbf{k} \sum_{i\omega_n} G^2_\mathbf{k}(i\omega_n)\partial_{k_x}\Sigma_\mathbf{k}(i\omega_n)$,
that associates $\Lambda^{\mathrm {conn}}(i\nu_0)$ with $\partial_{k_x}\Sigma_\mathbf{k}$ (see SM for further discussion). 
The conductivity is, however, determined by the slope,
$-\partial_{\nu}\mathrm{Re}\Lambda(i\nu)|_{\nu=0^+} = \sigma(\omega=0) = \sigma_{\mathrm dc}$,
and the contribution from $\Lambda^{\mathrm{conn}}$ is not small, but comparable to the bubble term.
The slope of the red line which corresponds to the DMFT $4\times 4$ - DMFT difference is small, reflecting the practically negligible finite-size effects in the bubble.

The shape of $\Lambda^{\mathrm{conn}}$ is difficult to reconstruct with analytical continuation from noisy data at the Matsubara frequencies (see SM), which we circumvented by using FTML.

Might the impact of vertex corrections change if larger systems are considered? The added
longer distance components of $\Lambda^\mathrm{conn}_\mathbf{r}$ could be sizeable, and even the short distance components might change due to improved diagrammatic content captured by the bigger cluster. We have performed the CTINT $8\times 8$ computation to address this question. In Fig.~\ref{fig:lambda_sigma}c we compare $\Lambda^\mathrm{conn}(i\nu_n)$ between $4\times 4$ and $8\times 8$ clusters (blue squares and black stars) and observe they are equal within the statistical error bars (about the size of the square symbol).
As for the longer distance components, we analyze the vertex corrections term as a function of real-space vector $\Lambda^\mathrm{conn}_\mathbf{r}(i\nu_n)$ and present the results in 
Fig.~\ref{fig:lambda_sigma}d. Indeed, the values drop rapidly with distance and the range of $\Lambda^\mathrm{conn}$ is clearly captured by the $4\times 4$ cluster.
Furthermore, the difference in the full $\Lambda$ between $4\times 4$
and $8\times 8$ clusters (purple crosses) appears to coincide with the
finite size effects in the bubble (red line/dots) obtained entirely
independently with DMFT.

Small finite-size effects are also indicated from a comparison of the frequency moments of FTLM $\sigma(\omega)$ in the high-$T$ limit with the exact values from Ref.~\onlinecite{huang18}, where we find an excellent agreement within $\lesssim 0.2$\% (see SM).

It is important to note that apart from reducing the dc resistivity, the vertex corrections have a characteristic effect on the frequency dependence of optical conductivity (see Fig.~\ref{fig:lambda_sigma}b and SM).
The high-frequency peak in $\sigma(\omega)$ obtained from DMFT is centered at precisely $\omega=U=2.5D$. This peak describes single-particle transitions between the Hubbard bands. The inclusion of vertex corrections brings about multi-particle excitations which move this peak towards lower frequencies, as noted previously in a slightly different context (see Refs.~\onlinecite{Cunningham2018,Gatti2007,Vidal2010}).

\emph{Conclusions.}
In the high-temperature $T\gtrsim t$, (doped) Mott insulator regime of
the Hubbard model, the single-particle self-energy is almost local, 
yet the vertex corrections to dc resistivity persist.
This finding applies to the optical lattice investigation in
Ref.~\onlinecite{brown18}, and explains why the DMFT results disagree 
with the experiment. On the other hand, we demonstrate that the long-distance
quantum fluctuations play a negligible role, and thus the $4\times 4$
isolated cluster becomes representative of 
the thermodynamic limit. The corresponding FTLM result is therefore
close to exact, and is an important benchmark for the
experiment in Ref.~\onlinecite{brown18} and future cold atoms
experiments.  

We cannot access with the same confidence the regime below $T\sim t$. 
Determinantal Quantum Monte Carlo algorithms in principle allow access to larger lattices and thus lower temperatures (see Ref.~\onlinecite{huang18}),
but the analytical continuation
presents a possible source of systematic error which is difficult to detect and estimate (see SM for a detailed analysis using the implementation of Maximum Entropy method taken from Ref.~\onlinecite{Levy_mem17}).
Our results highlight the need for developing real-frequency diagrammatic methods, like the one proposed recently in Ref.~\onlinecite{Taheridehkordi2019}.

Finally, our results suggest that proper account of the vertex corrections is needed \emph{at all temperatures}.
The discrepancies between the experimental observations and the DMFT, such as those observed 
in the case of hcp-Fe~\cite{Pourovskii2014} or in Sr$_2$RuO$_4$~\cite{deng16}
should not be interpreted only in terms of non-local correlations.
Very recently\cite{Kauch2019}, this conclusion has been shown to be valid even at much weaker coupling and in various other models.

\begin{acknowledgments}
We acknowledge useful discussions with V.~Dobrosavljevi\'c, A.~Georges, F.~Krien, and A.~M.~Tremblay, and contributions of A.~Vrani\'c and J.~Skolimowski
at early stage of this project. J.~K., R.~\v{Z}., and J.~M. are supported by
Slovenian Research Agency (ARRS) under Program P1-0044 and Project J1-7259.
J.~V and D.~T.~are supported by the Serbian Ministry of Education,
Science and Technological Development under Project No. ON171017.
Numerical calculations were partially performed
on the PARADOX supercomputing facility at the Scientific
Computing Laboratory of the Institute of Physics Belgrade.
The CTINT algorithm has been implemented using the TRIQS toolbox\cite{Parcollet2014}.
\end{acknowledgments}

\bibliography{refs.bib}

\begin{thebibliography}{54}
\bibitem [{com({\natexlab{a}})}]{comment3}%
  \BibitemOpen
  \href@noop {} {} ({\natexlab{a}}),\ \bibinfo {note} {{The irreducible vertex
  in the ph-channel is $\Gamma_{\mathrm{ph},\alpha\beta\gamma\delta} =\left.
  \frac{\delta \Sigma_{\delta\gamma}[G]}{\delta
  G_{\alpha\beta}}\right|_{G=G[G_0,\Sigma]}$ so even if a component of $\Sigma$
  is zero, it still depends on all components of the Green's function, and its
  derivative is not necessarily zero}}\BibitemShut {NoStop}%
\bibitem [{com({\natexlab{b}})}]{comment4}%
  \BibitemOpen
  \href@noop {} {} ({\natexlab{b}}),\ \bibinfo {note} {{Charge susceptibility
  is the uniform, zero-frequency component of the charge-charge correlation
  function $ \chi_\mathrm{c} \equiv \chi_{\mathbf{q}=0}(i\nu_n=0) \equiv \sum_i
  \int d\tau\; \Big\langle \big(n_i(\tau)-\langle n\rangle\big)\big( n_{i=0}(0)
  - \langle n \rangle\big) \Big\rangle $ where $n_i=n_{\uparrow
  i}+n_{\downarrow i}$ is the total charge operator. }}\BibitemShut {NoStop}%
\bibitem [{com({\natexlab{c}})}]{comment5}%
  \BibitemOpen
  \href@noop {} {} ({\natexlab{c}}),\ \bibinfo {note} {{The bubble part is
  expressed in terms of the full Green's function $G$ as $
  (\chi,\Lambda^{xx})^\mathrm{disc}_\mathbf{q}(i\nu_m) =-T\sum_{i\omega_n,
  \mathbf{k}, \sigma}
  v^2_\mathbf{k}G_{\sigma,\mathbf{k}+\mathbf{q}}(i\omega_n+i\nu_m)
  G_{\sigma,\mathbf{k}}(i\omega_n)$ with the vertex factor $v_\mathbf{k}=1$ in
  the case of $\chi$, and $v_\mathbf{k}=\frac{\partial
  \varepsilon_\mathbf{k}}{\partial k_x}=2t \sin k_x$ in the case of
  $\Lambda^{xx}$ on the square lattice. Here $\varepsilon_\mathbf{k}$ denotes
  the bare dispersion, $i\omega_n=i(2n+1)\pi T$ is the fermionic Matsubara
  frequency. }}\BibitemShut {NoStop}%
\bibitem [{com({\natexlab{d}})}]{comment2}%
  \BibitemOpen
  \href@noop {} {} ({\natexlab{d}}),\ \bibinfo {note} {in FTLM one can
  calculate the self-energy as well, but it is beyond the generality of our
  implementation.}\BibitemShut {Stop}%
\end{thebibliography}
\bibliographystyle{apsrev4-1}
\newpage
\begin{widetext}
\setcounter{page}{0}


\section*{Supplementary material}

# Supplemental Material: Conductivity in the square lattice Hubbard model at high temperatures: importance of vertex corrections

J. Vučičević,[1] J. Kokalj,[2,3] R. Žitko,[3,4] N. Wentzell,[5] D. Tanasković,[1] and J. Mravlje[3]

[1] Scientific Computing Laboratory, Center for the Study of Complex Systems,
Institute of Physics Belgrade, University of Belgrade, Pregrevica 118, 11080 Belgrade, Serbia

[2] University of Ljubljana, Faculty of Civil and Geodetic Engineering, Jamova 2, Ljubljana, Slovenia

[3] Jozef Stefan Institute, Jamova 39, SI-1000, Ljubljana, Slovenia

[4] University of Ljubljana, Faculty of Mathematics and Physics, Jadranska 19, Ljubljana, Slovenia

[5] Center for Computational Quantum Physics, Simons Foundation Flatiron Institute, New York, NY 10010 USA

Here we present a detailed analysis of the numerical results that we perform to disentangle the different contributions to the optical conductivity and identify the source of discrepancy between DMFT and FTLM. The analysis is performed on the imaginary axis where we can obtain the results from CTINT. Note that at high temperature, the Matsubara frequencies are far apart and the values of $\Lambda(i\nu_n)$ are insensitive to the details of $\sigma(\omega)$. We illustrate this in Fig. S1 where we show that, on the Matsubara axis, the FTLM and DMFT $\Lambda(i\nu_n)$ results are almost indistinguishable. However, the discrepancy is *not* below the level of noise in our numerics and we are able to reconstruct this difference from three different contributions, namely the finite-size effects, non-local self-energy effects and vertex corrections, all obtained independently using combinations of other methods. However, in the present context, we find the CTINT method useful only as a tool for benchmarking, since the analytical continuation from the imaginary to the real axis introduces a systematic error, and a precise $\sigma_{\mathrm{dc}}$ value is difficult to extract from $\Lambda(i\nu_n)$. In Section I we present our imaginary axis analysis of the results, and in Section II we discuss the difficulty of analytical continuation. Then, in Section III we benchmark our FTLM result against analytically computed frequency moments of the optical conductivity. In Section IV we discuss the details of the pole-broadening procedure used in FTLM.

## I. DETAILED BENCHMARK

In Fig. S2 we show the detailed comparison and cross-checks between the different methods in 12 doping-temperature $(p, T)$ points in the Hubbard model phase diagram at $U = 2.5D = 10t$. The continuous lines are obtained by the Hilbert transform from the real-axis to the continuous imaginary variable $\sigma(\omega) \rightarrow \Lambda(i\nu)$, and then taking the difference between the different methods, as written in the legend. The question we are addressing in the main text and that is considered in further detail here is the physical origin of the difference difference between DMFT and FTLM $4 \times 4$, presented

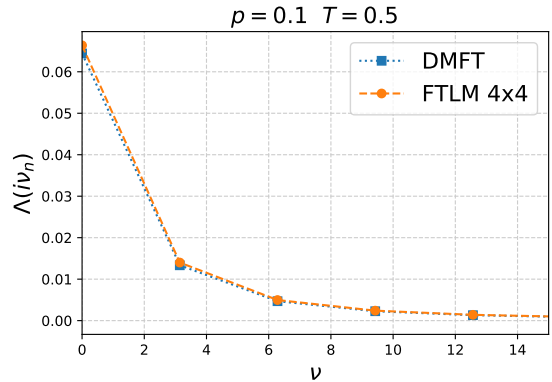

FIG. S1. Current-current correlation function $\Lambda(i\nu_n)$ in FTLM and DMFT (the dashed and dotted lines are guides to the eye).

on Fig. S2 by the orange lines.

We can readily inspect the effect of finite cluster size on the bubble $\Lambda^{\mathrm{disc}}$. This is given by the red line which presents the difference between DMFT $4 \times 4$ and DMFT. Red circles are obtained independently on the Matsubara axis without any analytical continuation, directly from DMFT data (DMFT here is performed with CTINT solver), and present an additional cross-check of our analytical continuation of the self-energy which was used to obtain $\sigma(\omega)$ in DMFT. We note that the statistical noise coming from CTINT in the single-site DMFT solution is very small, and the Padé analytical continuation of $\Sigma(i\omega_n)$ can be successfully performed. The optical conductivities agree closely (within few percent) between QMC and NRG solution.

We can also compare the red line with the difference between the full $\Lambda$ from CTINT $8 \times 8$ and $4 \times 4$ (purple crosses). The agreement is solid: it appears that the only difference between the $8 \times 8$ and $4 \times 4$ clusters is the finite-size effects in the bubble $\Lambda^{\mathrm{disc}}$, and that the finite-size effects disappear entirely already at cluster size $8 \times 8$. Note, however, that finite-size effects mostly pertain to the overall integral of $\sigma(\omega)$ (i.e. $\Lambda(i\nu = 0)$), and have little impact on $\sigma_{\mathrm{dc}}$.

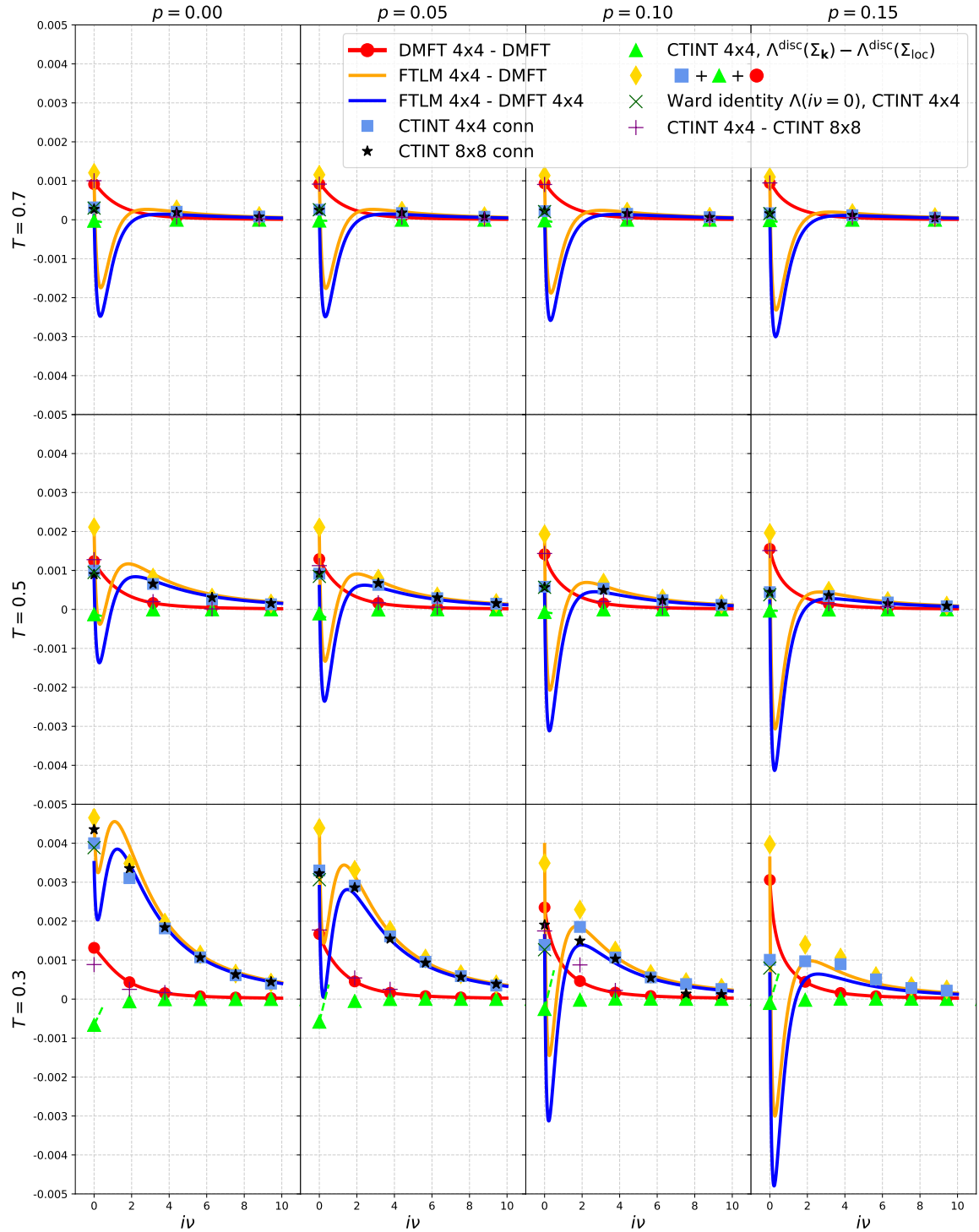

FIG. S2. Comparison of various parts of the current-current correlation functions $\Lambda(i\nu)$ on the imaginary axis (see text).

The blue line presents the difference between FTLM $4 \times 4$ and DMFT $4 \times 4$, which is by construction the orange line minus the red line, i.e. the difference between DMFT and FTLM $4 \times 4$, up to the finite-size effects in the bubble.

The blue squares and black stars are the vertex corrections $\Lambda^{\mathrm{conn}}$ as obtained from CTINT $4 \times 4$ and $8 \times 8$: at $T \geq 0.5D$ their agreement is excellent, and even at the lowest temperature it is likely within the statistical error bars of the method. At the lowest temperature there is some discrepancy but mostly due to increased statistical error in CTINT. The problem particularly pronounced at the biggest doping, where our CTINT $8 \times 8$ calculation suffers from the sign problem and failed to converge properly in the available computational time (384 `cpu*days` per point).

We have inspected also the self-energies and found excellent agreement between CTINT $8 \times 8$ and $4 \times 4$ (see Fig.3a in main text, other data not shown). We observe that the range of $\Sigma_{\mathbf{r}}$ is at most 2 lattice spacings, which means that the longer distance components that are captured by the $8 \times 8$ cluster are unlikely to have a measurable effect on any observable.

We cross check our results by calculating $\Lambda^{\mathrm{conn}}(i\nu = 0)$ from the Ward identity[1].

$$\Lambda^{\mathrm{conn}}(i\nu = 0) = -2T \sum_{\mathbf{k}} v_{\mathbf{k}} \sum_{i\omega_n} G_{\mathbf{k}}^2(i\omega_n) \partial_{k_x} \Sigma_{\mathbf{k}}(i\omega_n)$$

and present it using the dark-green cross. Here we have constructed $\Sigma_{\mathbf{k}}(i\omega_n)$ on the lattice ($64 \times 64$ grid Brillouin zone) using the Fourier transform of the short-distance $\Sigma_{\mathbf{r}}$ components available on the $4 \times 4$ cluster, which also allowed us to take the derivative analytically. Again, the agreement with the corresponding blue square and black star is within the roughly estimated statistical error of CTINT at all temperatures.

In most cases the blue line (difference between FTLM $4 \times 4$ and DMFT $4 \times 4$) passes through the blue squares (vertex correction from CTINT $4 \times 4$). However, at $i\nu = 0$ there appears to be a systematic deviation, and the blue line passes below the blue square. This we can link to the effect of non-local self-energy on the bubble which we calculate from the CTINT $4 \times 4$ results and present as green color triangles. Indeed, the green triangles are mostly negligible except at $\nu = 0$ where they are slightly negative.

We check our decomposition by summing the green triangles, blue squares and red circles, and comparing them to the orange line. Within statistical error bars, the total difference between FTLM$4 \times 4$ and DMFT appears to come from 1) finite-size effects in the bubble, 2) effects of non-local self-energy in the bubble and 3) vertex corrections.

Note, however, that the effects of non-local self-energy on the bubble are small and visible only at the lowest temperature, and related only to the overall integral of $\sigma(\omega)$, i.e. the kinetic energy. The only measurable effect on $\sigma(\omega = 0) = -\partial_\nu \Lambda(i\nu)|_{\nu \to 0^+}$ appears to come from the vertex corrections. We additionally cross check this by analytically continuing $\Sigma_{\mathbf{r} \neq 0}$ from CTINT $4 \times 4$ and using it together with DMFT $\Sigma_{\mathrm{loc}}(\omega)$ that we already have on the real-axis from NRG solver, to construct $\Sigma_{\mathbf{k}}(\omega)$ and calculate $\sigma^{\mathrm{disc}}(\omega = 0)$ . The difference from the pure DMFT result is negligible in relative terms except at $p = 0$ and lowest $T$ where $\sigma_{\mathrm{dc}}$ becomes very small. We present the corresponding slope in $\Lambda(i\nu)$ with green color dashed lines and see that it is much smaller that the slope of the blue line, and even in the opposite direction.

Based on the above analysis we conclude that at $T \gtrsim 0.3D$, finite-size effects and the effect of non-local self-energy on $\sigma_{\mathrm{dc}}^{\mathrm{disc}}$ become negligible, and that the vertex corrections $\sigma_{\mathrm{dc}}^{\mathrm{conn}}$ are already well converged with respect to the cluster size at the size $4 \times 4$. This builds confidence that our FTLM $4 \times 4$ is close to exact solution of the bulk Hubbard model.

## II. UNCERTAINTIES IN THE ANALYTICAL CONTINUATION OF $\Lambda(i\nu_n)$

In this section we thoroughly test the Maximum Entropy analytical continuation (MaxEnt) of the Matsubara current-current correlation function $\Lambda(i\nu_n) \to \sigma(\omega)$. We find that the result is strongly biased towards the model function used in MaxEnt continuation, and therefore discard the CTINT results for $\sigma(\omega)$ in favor of FTLM $4 \times 4$ which requires no analytical continuation.

In Fig. S3 we compare $\sigma(\omega)$ and $\Lambda(i\nu)$ between FTLM and DMFT. As a function of continuous imaginary variable, $\Lambda(i\nu)$ is displayed by a line, and the Matsubara frequencies are indicated with crosses. Note that only the values at the Matsubara frequencies $\Lambda(i\nu_n)$ serve as the input for MaxEnt. We see that most of the difference between FTLM and DMFT is encoded between the first two Matsubara frequencies in $\Lambda(i\nu)$. In particular, the dc conductivity is given by $\sigma_{\mathrm{dc}} = -\partial_\nu \Lambda(i\nu)|_{\nu \to 0^+}$, which is hard to estimate based on $\Lambda(i\nu_n)$. Although there is a one-to-one correspondence between any given function on the real axis and its Hilbert transform on the imaginary axis, any amount of noise in $\Lambda(i\nu_n)$ and a truncation of Matsubara frequencies is likely to lead to loss of critical information necessary to distinguish between two similar $\sigma(\omega)$.

Fig. S4 shows the optical conductivity obtained by the analytical continuation of the current-current correlation function $\Lambda(i\nu_n)$ from CTINT. We use the implementation of the Maximum Entropy method from Ref. 2. We put the error bar $d\Lambda(i\nu_n) = 10^{-4}$. Below this value the MaxEnt $\sigma(\omega)$ starts to acquire noisy and

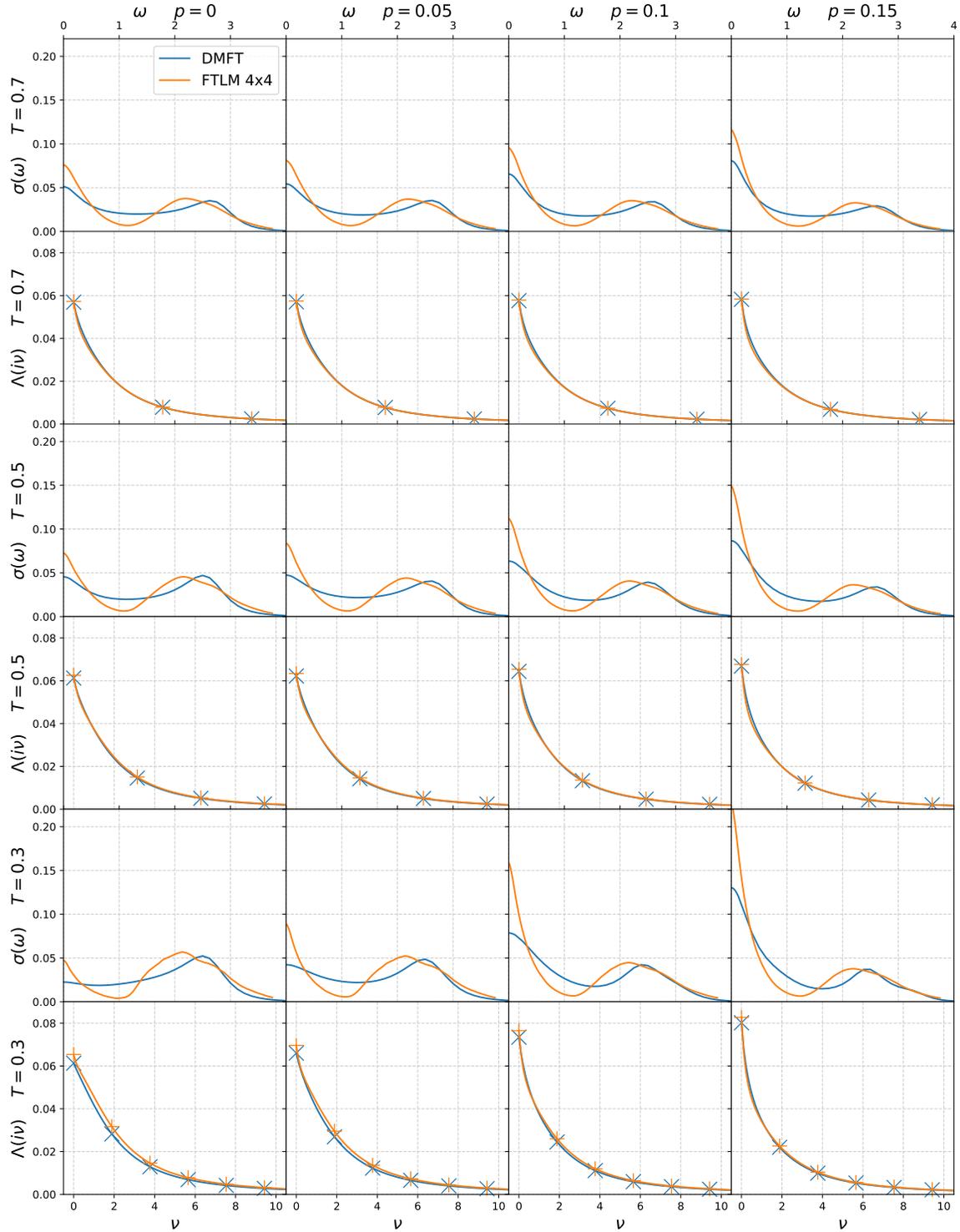

FIG. S3. Optical conductivity and current-current correlation function (see text).

manifestly wrong features, due to overfitting. This value of $10^{-4}$ also agrees with the deviation in $\Lambda(i\nu_n)$ between CTINT $4 \times 4$ and FTLM, attributed to the statistical noise in CTINT. We perform annealing similar to Ref. [3]: we apply MaxEnt at temperature $T = 0.5D$, using either FTLM (left column) or DMFT (middle column) $\sigma(\omega)$ at $T = 0.7$ as the default model. MaxEnt is then done at $T = 0.3D$, using the result of previous MaxEnt as the default model. The right column in Fig. S4 shows the resulting dc resistivities.

We see that the result of the analytical continuation strongly depends on the initial model function at high temperature. Furthermore, when the initial model is given by FTLM, the result at $T = 0.3$ still tends to deviate towards the DMFT solution. The reason for this is that the Drude-like peak in DMFT is broader than in FTLM, and the MaxEnt generally tends to make the spectrum smoother. This means that even with the correct default model at the highest temperature, the error bar introduced by annealing can easily erase any information about the vertex corrections and produce a result comparable to just the bubble contribution that one can safely obtain from DMFT(NRG). When the initial default model is taken to be DMFT, the error bar goes up to 50 percent, and the results typically resemble the DMFT solution.

Instead of choosing as the default model the FTLM results, which are computationally expensive to obtain (around one month on 32 cores with 80 GB of RAM for single choice of boundary conditions), it may appear reasonable to try and start the annealing using the high-$T$ expansion[4] result at the highest temperature. However, as shown in Ref. [3] even high-$T$ expansion is not trivial to calculate, and can only yield $\sigma(t)$ results up to $t \approx 1$ ($t$ here is real time). In Fig. S5 we illustrate how the short time conductivity holds little information about $\sigma_{\rm dc}$ as $\sigma_{\rm dc} \sim \int dt \, {\rm Re} \sigma(t)$. The error made in the high-$T$ expansion then propagates in MaxEnt, and can lead to wrong results.

Finally, it should be noted that with increasing temperature, Matsubara frequencies spread out, leaving less and less information to be extracted from even a slightly noisy $\Lambda(i\nu_n)$. We conclude that doing MaxEnt on CTINT $8 \times 8$ even with the corresponding FTLM $4 \times 4$ default model would not bring any information other than what is already contained in FTLM. Our analysis highlights the importance of developing methods that calculate the current-current correlation function directly on the real frequency axis.

## III. COMPARISON WITH THE MOMENTS FROM THE HIGH-TEMPERATURE EXPANSION

In the high-$T$ limit with $\sigma(\omega) \propto 1/T$, the frequency moments $\mu_k = \frac{1}{2\pi} \int_{-\infty}^{\infty} \sigma(\omega) \omega^k d\omega$ can be calculated reliably or even analytically[3,4] as the expectation values of certain commutators between the Hamiltonian and the current operator. Despite the difficult to reconstruct $\sigma(\omega)$, and in particular $\sigma_{\rm dc}$ from such moments with high confidence[3], the moments still provide a firm test of the numerical approaches.

By using the real frequency $\sigma(\omega)$ obtained with FTLM, we calculate frequency moments in the high-$T$ limit for $U = 1.5D$ and $p = 0.2$. Such moments can be compared to the exact values reported in Ref. [3]. We find that our FTLM moments $\mu_k$ for $k = 0 - 8$, which have main contributions from $\sigma(\omega)$ in the regime $|\omega| \lesssim 4D$ (i.e. up to $\omega$ about $2D$ above the upper edge of the Hubbard band), deviate from the exact moments by $\lesssim 0.2\%$. Some lower moments show even smaller deviation (see Table S1), which suggest FTLM correctly reproduces high-$T$ behavior with small finite size effects. Our higher moments ($k \gtrsim 10$) show systematic larger deviation from the exact results due to high frequency cutoff at $\omega > 5D$ in our FTLM results.

$T \to \infty$ values of the FTLM moments are obtained by fitting $T$ dependence of $2T\mu_k$ to $a + b/T^2$ in the temperature range between $5D$ and $10D$. The numerical uncertainties given in brackets in the Table S1 are obtained as a standard deviation in the fitting procedure.

| k | $2T\mu_k$ (exact) | $2T\mu_k$ (FTLM) |
|---|---|---|
| 0 | 0.96 | 0.96001(9) |
| 2 | 16.5888 | 16.554(4) |
| 4 | 879.206 | 879.4(2) |
| 6 | 71350.4 | 71525(20) |
| 8 | $7.95719 \cdot 10^6$ | $7.963(2) \cdot 10^6$ |

TABLE S1. Exact frequency moments $2T\mu_k$ taken from Ref. [3] and the moments from integrating FTLM $\sigma(\omega)$ (here the units of $t = D/4 = 1$ are used). The numbers in the brackets are estimates of numerical uncertainty for the last digits. Small deviations of FTLM moments from exact values suggest small finite size effects in the high-$T$ limit.

## IV. BROADENING IN FTLM

Optical conductivity calculated with FTLM on a finite cluster is strictly a set of delta functions in frequency space. The number of such delta functions grows with the number of many-body states, leading to a high density for the used cluster sizes. Still, the delta functions

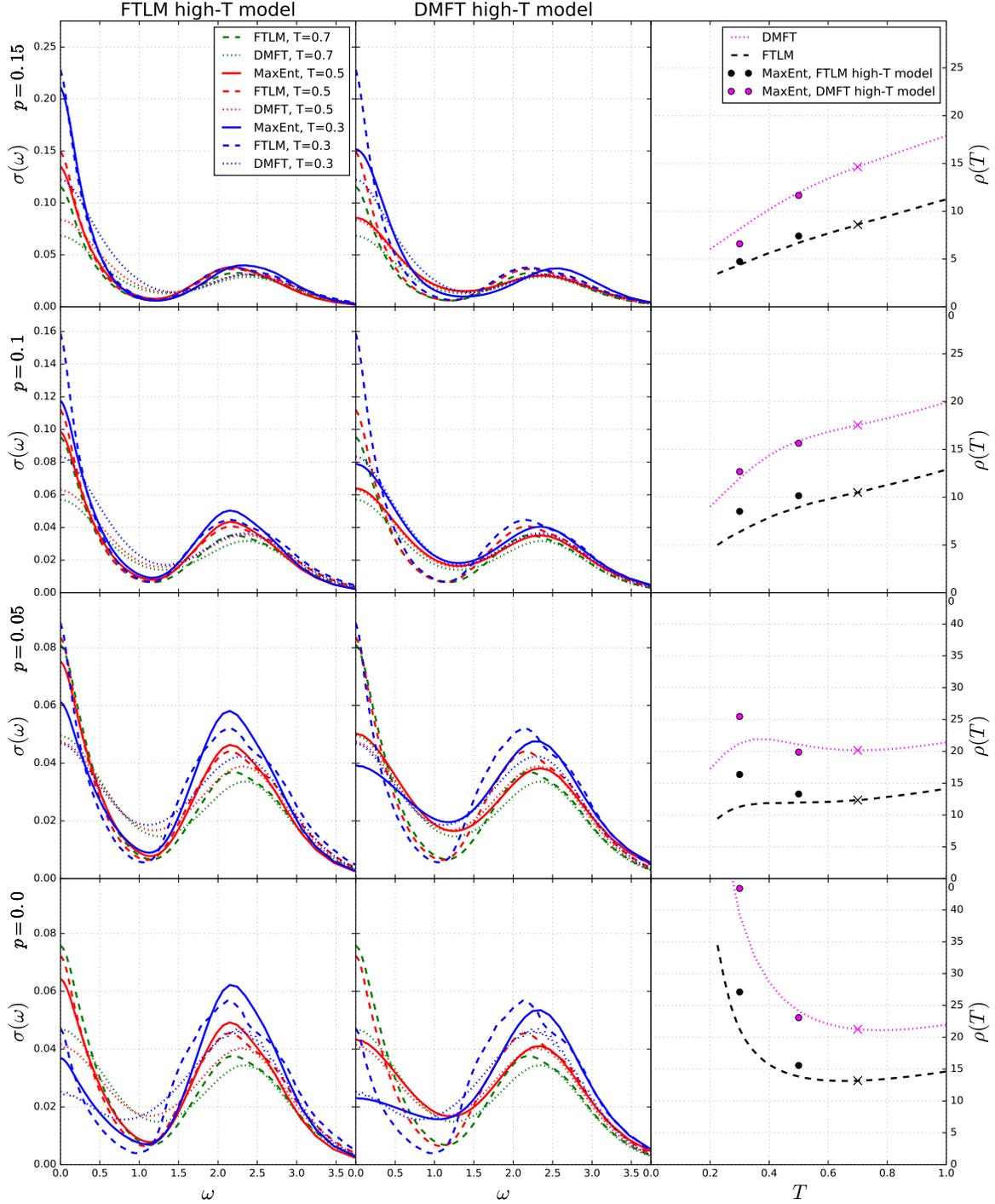

FIG. S4. Optical conductivity obtained by MaxEnt analytical continuation of the CTINT $\Lambda(i\nu_n)$ (solid lines in the first and the second column). The annealing method is used, where the initial model function, used for MaxEnt at $T = 0.5$, is the FTLM (first column), and DMFT (second column) at $T = 0.7$. At $T = 0.3$ the model function is the MaxEnt result from $T = 0.5$. The dashed (dotted) lines are FTLM (DMFT) data. The right column shows the MaxEnt resistivities, $\rho = 1/\sigma(\omega = 0)$, in comparison with FTLM and DMFT $\rho(T)$ curves. The four rows correspond to different doping levels $p = 0.15, 0.1, 0.05, 0$. The crosses are the dc resistivity corresponding to the initial model function.

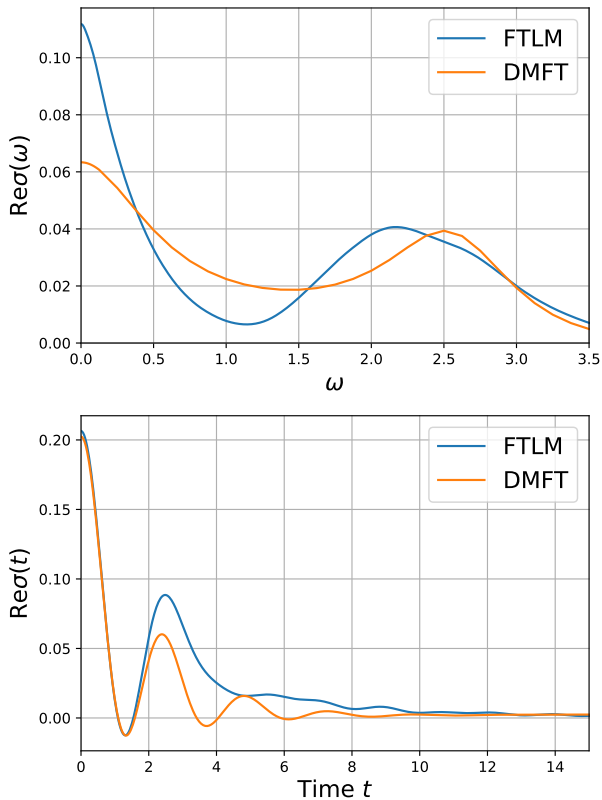

FIG. S5. Optical conductivity and its Fourier transform at $p = 0.1$, $T = 0.5D$. The dc conductivity has contributions from up to $t \sim 10$. DMFT and FTLM practically coincide at $t < 1$.

need to be broadened to get a smooth spectra, representative of the thermodynamic limit. The value of the broadening needs to be appropriate: sufficiently large to remove the finite-size artifacts, but not large enough to over-broaden the real features of the spectrum[5,6].

In our case we use Gaussian broadening, with the broadening parameter chosen as the parameter for which $\sigma_{dc}$ is not changing or shows smallest change with broadening, a choice to which we refer as the optimal one. See Fig. S6. This prescription works also for finite and high frequencies, where the delta functions are denser

and the spectra are smooth even with smaller broadening parameter. The used optimal broadening parameter is substantially smaller than the width of the Drude peak and we estimate the broadening uncertainty of $\rho_{dc}$ within FTLM to be below 10%.

It is worth noting that that with increasing broadening the $\sigma_{dc}$ drops monotonically. Since in all cases $\sigma_{dc}$ in DMFT is lower, there must always be a certain broadening level that reproduces the DMFT result for $\sigma_{dc}$, but not simultaneously $\sigma(\omega)$ at all frequencies. We have checked that the broadening level needed to reproduce $\sigma_{dc}$ from DMFT is about 10 times the optimal one, and becomes comparable to the width of the Drude peak. This choice of broadening leads to sever modification in the shape of $\sigma(\omega)$, especially of the high-frequency peak which is otherwise well determined already by a fine binning of delta functions or with a tiny broadening. Therefore, we exclude such large broadening from consideration.

Finally, we note that for the calculation of $\Lambda(i\nu_n)$ from $\sigma(\omega)$ obtained by FTLM with Hilbert transform, Eq. (3) in the main text, no broadening is needed due to integration and that even if the broadened $\sigma(\omega)$ is used, $\Lambda(i\nu_n)$ change by the order of $10^{-5}$, which is smaller than the symbol size in Fig. 3 (main text) and in Fig. S2 and is also below the CTINT noise level.

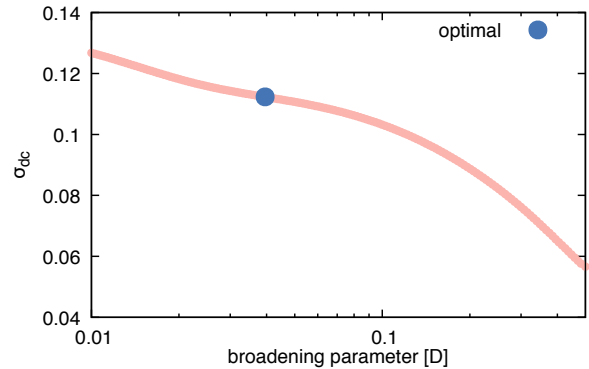

FIG. S6. Representative dependence of $\sigma_{dc}$ from FTLM on the broadening parameter and the optimal parameter according to the minimal change of $\sigma_{dc}$ with broadening. Data are for $p = 0.1$ and $T = 0.5D$.


\end{widetext}
\end{document}